1# Frank McClean and the Ferncliffe Observatory at Tunbridge Wells

**Jeremy Shears**

Introduction

My abiding memories of the late RAS Librarian Peter Hingley include the wonderfully entertaining conversations we had based on his encyclopaedic astronomical knowledge. In addition, there appeared to be no part of the country with which Peter was not acquainted and he generally had an anecdote to tell about each. For example, noting that I live near Chester, he told me about his visits to the City as well as the hours he spent in the library of Hawarden Castle, just across the Welsh border in Flintshire. Similarly when I mentioned that

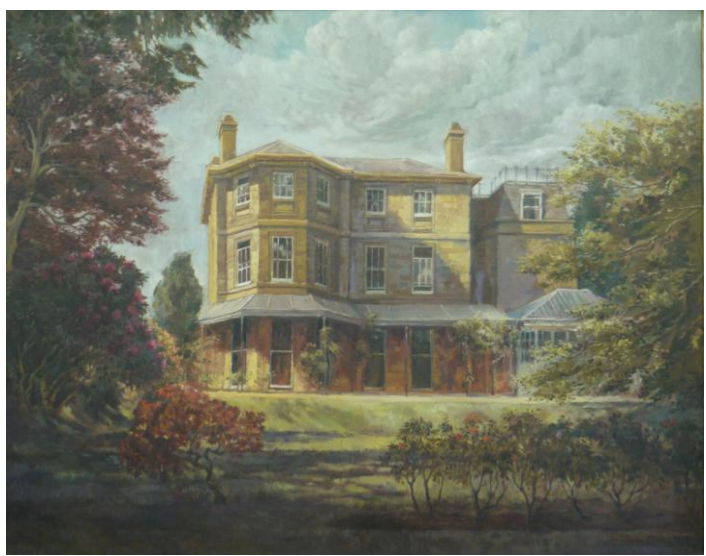

Figure 1: Ferncliffe, Tunbridge Wells. Courtesy of Mrs V. Hawkins, Concord College, Shropshire (7)

I had grown up in Kent, he told me about the Ferncliffe Observatory at Tunbridge Wells, which had been built by the pioneering spectroscopist, Frank McClean (1837-1904), in the 1870s. This came as something of a surprise as I had no idea that Tunbridge Wells was once home to an important astronomical research centre. Moreover, it transpired that McClean's residence (Figure 1) was located on the Pembury Road, one of the main routes into the town, and therefore I must have unwittingly driven past it dozens of times over the years. Peter kindly gave me copies of some photographs of McClean's observatory from the RAS archives along with permission to publish them.

This note provides a brief overview of McClean's life and some details of his Ferncliffe Observatory to accompany the photographs, which had been given to the RAS by McClean himself. Further information may be obtained from his several obituaries (1).

**Early life, family and career**

Born in Down, Ireland, on 13 November 1837 Frank McClean was the only son of the wealthy civil engineer John Robinson McClean, CB, FRS (1813-1873). He attended Westminster School, going on to Glasgow University, where he studied under Professor William Thomson (1824-1907; later Lord Kelvin), and then to Trinity College Cambridge, graduating as 27$^{th}$ Wrangler in the Mathematics Tripos in 1859. He joined his father's firm of McClean & Stileman, becoming a partner in 1862. He worked for some time as an engineer in the Barrow dockyards (2).

*Published in the Newsletter of the British Astronomical Association's Historical Section*

McClean married Ellen Greg (1842-1909), of Escowbeck, Lancashire, in 1865. To what extent this union increased his personal fortune, or whether it was solely from his father's firm, he was nevertheless in a position to retire in 1870 and to devote the rest of his life to his scientific and artistic interests. The couple settled at Ferncliffe, an imposing property just outside Tunbridge Wells, in 1871 (3). They had three sons and two daughters. One son, Lieutenant-Colonel Sir Francis Kennedy McClean (1876-1955), also became a keen amateur astronomer under his father's influence. He would also gain public acclaim as a pioneer aviator. In 1911 he became a celebrity overnight when he flew a Short Brothers aeroplane between the towers of Tower Bridge in London.

**The Ferncliffe Observatory**

McClean's first experiments at Ferncliffe were in connexion with electricity, but in 1875 he built an observatory in which he installed a 15-inch (38 cm) With-Browning reflector and a 6-inch (15 cm) Merz refractor (Figure 2) (4). The observatory itself (Figure 3) was essentially a wooden apex shed just large enough to contain the instruments, with two halves joined by a hinge that allowed them to be swung apart for observation. The shed was mounted on an elevated wooden deck, itself supported on brick foundations. The German equatorial was mounted on two brick plinths which passed through the decking into the ground below.

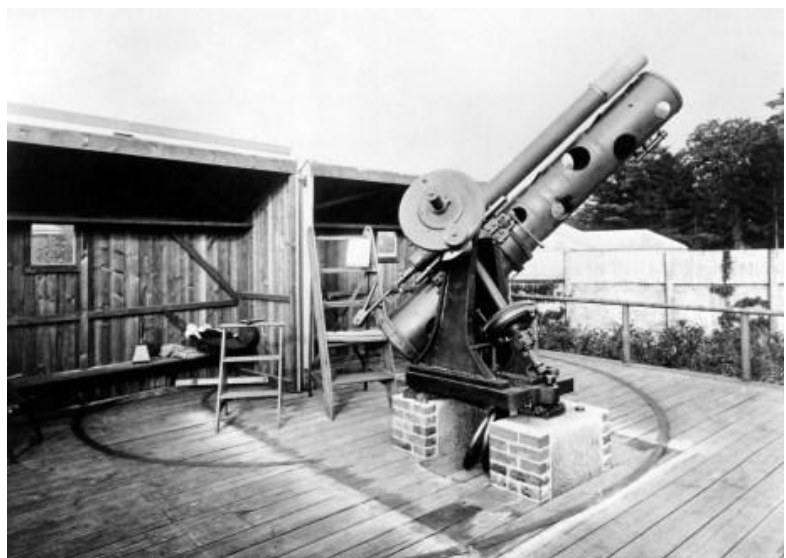

Figure 2: The 15-inch With-Browning reflector and 6-inch Merz refractor at Ferncliffe, ca. 1879. Courtesy RAS.

McClean soon became interested in astronomical spectroscopy and in doing so developed a new design of spectroscope, which was named after him. Existing spectroscopes were difficult to use because it was necessary to align the star with a tiny slit. McClean's innovation was to use a concave cylindrical lens to overcome this. In practise this meant that once a star was lined up with a telescope, the eyepiece could be removed and the spectroscope inserted. This allowed the device to be used on a wider range of telescopes, and became instantly popular with amateur and professional astronomers. The patented instrument design was commercialised by the firm of John Browning of London (Figure 4).

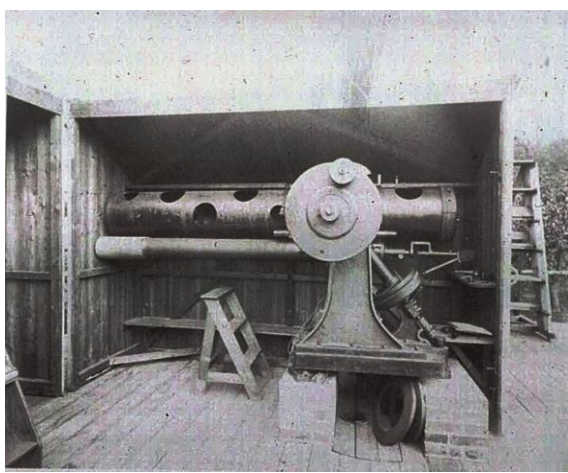
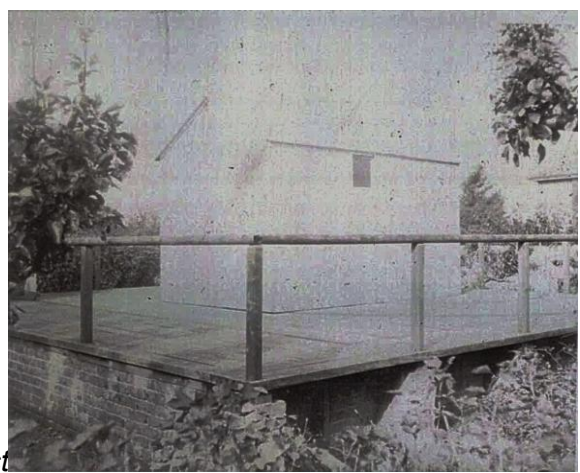

Figure 3: The Ferncliffe Observatory. Courtesy RAS

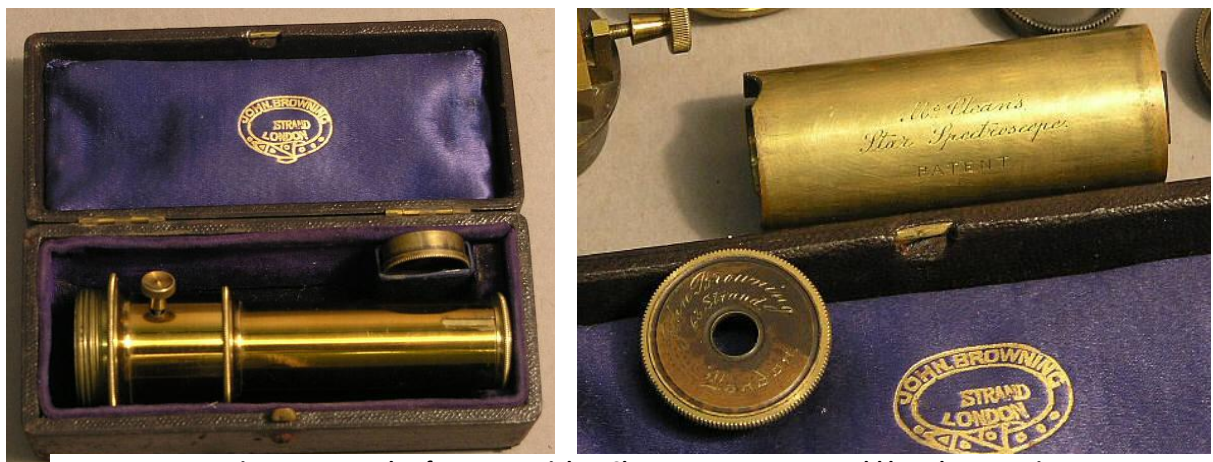

Figure 4: Example of a commercial McClean spectroscope, as sold by John Browning

I have no information about what became of the Ferncliffe instruments and would be delighted to hear from anyone who can shed any light on the matter.

**Life after Ferncliffe**

In 1884 the McClean family moved from Ferncliffe to a new home, Rusthall House, on the other side of Tunbridge Wells. There, in addition to a polar heliostat that was located in the roof of the house, McClean built a new observatory containing twin refractors: a 10-inch (25 cm) for visual use and a 12-inch (30 cm) photographic, the latter with an objective prism, which are now located at the Norman Lockyer Observatory at Sidmouth. These telescopes are probably the one depicted on the rather charming bookplate used in McClean's library books (Figure 5). With them, he conducted a spectroscopic survey of some 160 stars in the northern sky. In 1897 he went to the Cape of Good Hope to extend the survey to the southern hemisphere and in the process identified oxygen in the spectrum of β Crucis and other stars. The work earned him the Gold Medal of the RAS in 1899.

McClean was a generous benefactor, presenting the Cape of Good Hope Observatory with the

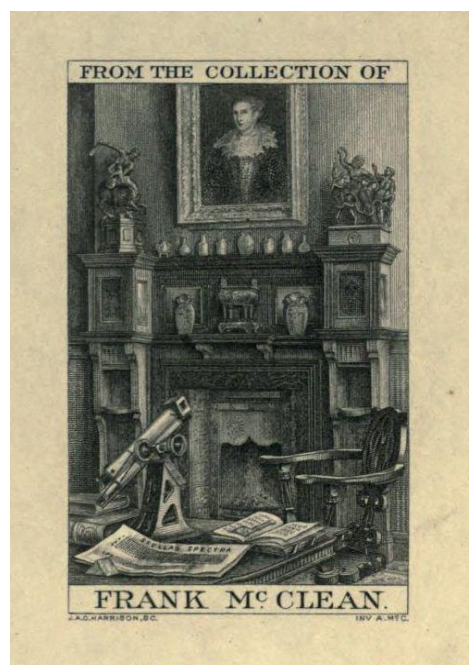

Figure 5: Bookplate of Frank McClean; note the model telescope and the paper on "Stellar Spectra" on the table. He collected objets d'art, some of which can also be seen (6)

*Published in the Newsletter of the British Astronomical Association's Historical Section*

Victoria 24-inch (61 cm) photographic telescope and an 18-inch (46 cm) visual instrument. He founded three Isaac Newton studentships in astronomy at Cambridge University. He also left bequests to Birmingham and Cambridge Universities, the Royal Society, the RAS and the Royal Institution.

McClean was an original member of the BAA and served on Council between 1900 and 1902. He was elected FRAS in 1887 (5), FRS in 1895 and was awarded an honorary LLD degree by Glasgow University in 1894. He died in Brussels on 8 November 1904 in his 67$^{th}$ year.

**Address:** "Pemberton", School Lane, Bunbury, Tarporley, Cheshire, CW6 9NR, UK [bunburyobservatory@hotmail.com]

**Notes and references**

1. *Obituaries: Turner H.H., MNRAS, 65, 338-342 (1905); Newall H.F., Observatory, 27, 449-450 (1904); and JBAA, 15(1), 47 (1904).*

2. *McClean was also resident-engineer for the Furness and Midland Joint Railway.*

3. *McClean also retained a residence in South Kensington, London: first at 21 Onslow Square and then at Onslow Gardens.*

4. *The year in which the observatory was built remains uncertain. Whilst the MNRAS obituary says it was 1875, the RAS photographs themselves are annotated "Ferncliffe Obs., Tunbridge Wells, Kent. Erected 1879 by F McClean who gave the photos".*

5. *McClean served on the RAS Council from 1891 almost continuously until his death.*

6. *McClean's valuable collection of objets d'art from around the world, including glassware, ancient stone carvings and medieval manuscripts were left to the Fitzwilliam Museum, Cambridge, upon his death. He sometimes bought items under the pseudonym "Money".*

7. *Concord College was housed in Ferncliffe from shortly after its founding in 1949 until 1973, when it moved to Shropshire. The painting was commissioned by the College at about the time of the move and it presently hangs in the Staff Room.*